\documentclass[12pt]{article}
\usepackage{epsf}
\usepackage{subfigure}
\title{Enhanced electroweak radiative corrections in SUSY
and precision data}
\author{I.V.~Gaidaenko, A.V.~Novikov, V.A.~Novikov \\
ITEP, Moscow, Russia \\
A.N.~Rozanov \\
CPPM, IN2P3, CNRS, France and ITEP, Moscow, Russia \\
M.I.~Vysotsky \\
ITEP, Moscow, Russia and INFN, Sezione di Ferrara, Italy}
\date{\it Submitted to Physics Reports}
\begin{document}
\maketitle
\def\la{\mathrel{\mathpalette\fun <}}
\def\ga{\mathrel{\mathpalette\fun >}}
\def\fun#1#2{\lower3.6pt\vbox{\baselineskip0pt\lineskip.9pt
\ialign{$\mathsurround=0pt#1\hfil##\hfil$\crcr#2\crcr\sim\crcr}}}
\newcommand{\nuanu}[1]{
 \stackrel{{\scriptscriptstyle (-)}}
                          {\nu}_{\hspace{-3pt}#1}}
\newcommand{\subn}[1]{\mbox{\scriptsize #1}}
\newcommand{\thetaW}{\theta_{\subn{W}}}
\newcommand{\sW}{s_{\subn{W}}}
\newcommand{\cW}{c_{\subn{W}}}
\newcommand{\cu}{c_{\subn{u}}}
\newcommand{\su}{s_{\subn{u}}}
\newcommand{\mZ}{m_{\subn{Z}}}
\newcommand{\mW}{m_{\subn{W}}}
\newcommand{\mWZ}{m_{\subn{W,Z}}}
\newcommand{\SigmaW}{\Sigma_{\subn{W}}}
\newcommand{\SigmaV}{\Sigma_{\subn{V}}}
\newcommand{\SigmapWZ}{\Sigma_{\subn{W,Z}}'}
\newcommand{\mH}{m_{\subn{H}}}
\newcommand{\mh}{m_{\subn{h}}}
\newcommand{\mA}{m_{\subn{A}}}
\newcommand{\mt}{m_{\subn{t}}}
\newcommand{\mSUSY}{m_{\subn{SUSY}}}
\newcommand{\ndf}{n_{\subn{d.o.f.}}}
\newcommand{\Ee}{E_{e}}
\newcommand{\pt}{p_{\subn{t}}}
\newcommand{\ppt}{p_{\subn{t}}^2}
\newcommand{\gV}{g_{\subn{V}}}
\newcommand{\gA}{g_{\subn{A}}}
\newcommand{\Vq}{\mbox{\rm V}_{\subn{q}}}
\newcommand{\Aq}{\mbox{\rm A}_{\subn{q}}}
\newcommand{\iq}{i_{\subn{q}}}
\newcommand{\gVq}{g_{\subn{$\Vq$}}}
\newcommand{\gAq}{g_{\subn{$\Aq$}}}
\newcommand{\RVq}{R_{\subn{$\Vq$}}}
\newcommand{\RAq}{R_{\subn{$\Aq$}}}
\newcommand{\Riq}{R_{\subn{$\iq$}}}
\newcommand{\gVe}{g_{\subn{V}}^{e}}
\newcommand{\gAe}{g_{\subn{A}}^{e}}
\newcommand{\gR}{g_{\subn{R}}}
\newcommand{\gL}{g_{\subn{L}}}
\newcommand{\gRL}{g_{\subn{R(L)}}}
\newcommand{\gLR}{g_{\subn{L(R)}}}
\newcommand{\thetaLR}{\theta_{\subn{LR}}}
\newcommand{\dSUSYLR}{\delta^{\subn{LR}}_{\subn{SUSY}}}
\newcommand{\dSUSY}{\delta_{\subn{SUSY}}}
\newcommand{\Eth}{\Ee\theta_{e}^2}
\newcommand{\nue}{\nu_{e}}
\newcommand{\anue}{\bar\nu_{e}}
\newcommand{\numu}{\nu_{\mu}}
\newcommand{\anumu}{\bar\nu_{\mu}}
\newcommand{\nutau}{\nu_{\tau}}
\newcommand{\anumue}{\bar\nu_{\mu}e}
\newcommand{\gn}{g^{\nu}}
\newcommand{\gnumu}{g^{\nu_{\mu}}}
\newcommand{\gnue}{g^{\nu_{e}}}
\newcommand{\gnutau}{g^{\nu_{tau}}}
\newcommand{\sbot}{\tilde{\mbox{\rm b}}}
\newcommand{\stot}{\tilde{\mbox{\rm t}}}
\newcommand{\stL}{\tilde{\mbox{\rm t}}_{\subn{L}}}
\newcommand{\stR}{\tilde{\mbox{\rm t}}_{\subn{R}}}
\newcommand{\stone}{\tilde{\mbox{\rm t}}_1}
\newcommand{\sttwo}{\tilde{\mbox{\rm t}}_2}
\newcommand{\sbL}{\sbot_{\subn{L}}}
\newcommand{\sbR}{\sbot_{\subn{R}}}
\newcommand{\msg}{m_{\subn{$\tilde{\mbox{\rm g}}$}}}
\newcommand{\msq}{m_{\subn{$\tilde{\mbox{\rm q}}$}}}
\newcommand{\mstL}{m_{\subn{\mbox{$\stL$}}}}
\newcommand{\mstR}{m_{\subn{\mbox{$\stR$}}}}
\newcommand{\msb}{m_{\subn{\mbox{$\sbot$}}}}
\newcommand{\msbL}{m_{\subn{\mbox{$\sbL$}}}}
\newcommand{\msbR}{m_{\subn{\mbox{$\sbR$}}}}
\newcommand{\Vm}{V_{\subn{m}}}
\newcommand{\VR}{V_{\subn{R}}}
\newcommand{\VA}{V_{\subn{A}}}
\newcommand{\YL}{Y_{\subn{L}}}
\newcommand{\alphaW}{\alpha_{\subn{W}}}
\newcommand{\alphashat}{\hat{\alpha}_{\subn{s}}}
\newcommand{\alphabar}{\bar{\alpha}}

\newcommand{\unit}[1]{\, {\rm #1}}
\newcommand{\laeq}{\raisebox{-0.3ex}
                   {\footnotesize $\stackrel{\raisebox{-0.8ex}
                                              {$\textstyle <$}}{\sim}$}}
\newcommand{\diff}[2]{\frac{{\rm d}\,#1}{{\rm d}\,#2}}
\newcommand{\anu}{\bar\nu}
\newcommand{\andd}{\hspace*{1em} {\rm and}\ \hspace{1em}}
\newcommand{\with}{\hspace*{3em} {\rm with}\ \hspace{1em}}
\newcommand{\where}{\hspace*{3em} {\rm where} \hspace{1em}}

\begin{abstract}

Enhanced radiative corrections  generated in SUSY
extensions of the Standard Model spoil the fit of the precision
data (Z-boson decay parameters and W-boson mass).
This negative effect is washed out for heavy enough squarks,
because of the decoupling property of SUSY models.
We find that even for light squarks the enhanced radiative corrections can be
small. In this case substantial $\stL \stR$ mixing is
necessary.

\end{abstract}

\newpage

A number of predictions of the Standard Model are now tested with an
accuracy of the order of 0.1\%. There are two (more or less) free parameters
in the Standard Model: the value of the strong coupling constant
 $\alphashat(\mZ)$
and the mass of the Higgs boson $\mH$.
 Making a fit of the latest set of
precision data reported in \cite{1} (Z-boson decay parameters, W-boson
and top-quark masses), we obtain:
\begin{equation}
\mH = (71^{+82}_{-43}) \; \mbox{\rm GeV} \;\; ,
\label{1}
\end{equation}
\begin{equation}
\alphashat(\mZ) = 0.119 \pm 0.003 \;\; ,
\label{2}
\end{equation}
\begin{equation}
\chi^2/ \ndf = 15.0/14 \;\; .
\label{3}
\end{equation}

The quality of this fit is very good, which imposes strict constraints on the
possible extensions of the Standard Model.

In the SUSY extensions a lot of new particles are introduced.
 Their contributions
to electroweak observables come through loops and
for $\mSUSY > \mW$ are of the order of
$\alphaW$ $(\mW/\mSUSY)^2$,
 where $\mSUSY$ characterises the mass scale of the
new particles. These contributions were calculated and
analysed in number of papers \cite{2}-\cite{5}.

A large violation of SU(2) symmetry in the third family of squarks by a large
value of $\mt \approx 175$ GeV leads to an enhancement of the
corresponding oblique corrections by the factor
 $(\mt/\mW)^4 \approx 16$ to be
 compared with the numerous terms of the order of
 $\alphaW(\mW/\mSUSY)^2$.
(The presence of terms $\sim \mt^4$
 in SUSY models was recognised long ago
\cite{6}.) Inspired by this fact, calculations of the enhanced corrections to
the functions $\VA$, $\VR$ and $\Vm$
 were made \cite{7}. The functions $V_i$
describe electroweak radiative corrections to Z-boson couplings to
leptons $\gA$ and $\gV/\gA$, and to
 W-boson mass \cite{8}. To calculate
these enhanced terms we expanded the polarisation operators of
the vector bosons $\SigmaV (k^2)$ at
$k^2 =0$. The terms enhanced as
 $\mt^4/\mW^2 \mSUSY^2$ come from
$\SigmaW(0)$, while those enhanced as
 $\mt^2/\mSUSY^2$ come from
$\SigmapWZ (0)$.
  The higher-order derivatives of self-energies are suppressed
as $(\mWZ/\mSUSY)^2$, and are therefore
not taken into account.

In the present paper the influence of these new terms on the precision data fit
will be analysed. To begin with, we should expand the analysis of \cite{7} and
take into account the main SUSY corrections to hadronic Z-decays as well.
Vertices with gluino exchange generate (potentially) large corrections of
the order of $\alphashat(\mZ/\mSUSY)^2$
 in the limit $\mSUSY > \mZ$. For
hadronic Z decays we use the following expression for the width \cite{9}:
\begin{equation}
\Gamma_{\mbox{\rm q}}
 = \Gamma_{\mbox{\rm Z}\to \mbox{\rm q}\bar{\mbox{\rm q}}}
 = 12[\gAq^2 \RAq +
      \gVq^2 \RVq ]
\Gamma_0 \;\;,
\label{4}
\end{equation}
where $\Gamma_0=\frac{1}{24\sqrt{2}\pi}G_{\mu}\mZ^3$.

Corrections induced by gluino exchanges lead to the
following SUSY shifts of the factors $\Riq$ \cite{10}:
\begin{equation}
\delta \RVq =
\delta \RAq =
 1+ \frac{\alphashat (\mZ)}{\pi}
\Delta_1 \;\; ,
\label{5}
\end{equation}
\begin{equation}
\Delta_1 = -\frac{4}{3}\int\limits_0^1 dz_1 \int\limits_0^{1-z_1} dz_2 \ln
\biggl[ 1- \frac{xyz_1 z_2}{x+(z_1 +z_2)(y-x)} \biggr] \;\; ,
\label{6}
\end{equation}
where $x =(\mZ / \msq )^2$,
 $y=(\mZ / \msg )^2$.
We take these strong SUSY corrections into account in our analysis. The
weak SUSY
corrections to Z decays into hadrons are taken into account by the corrections
to the functions $\VA$ and $\VR$ calculated in \cite{7}.

The stop exchange contributes to the vertex corrections to the
 $\mbox{\rm Z} \to \mbox{\rm b}\bar{\mbox{\rm b}}$
decay amplitude; but since there are no terms enhanced as $(\mt / \mW )^4$
\cite{11}, we will not take the corresponding corrections into account in
the present paper.

Let us start the discussion of the SUSY corrections to the functions $V_i$
from the description of the stop sector of the theory.
The $\stL \stR$ mass matrix has the following form:
\begin{equation}
\left(
\begin{array}{cc}
m^2_{\stL} & \mt A'_{\mbox{\rm t}} \\
\mt A'_{\mbox{\rm t}} & m^2_{\stR}
\end{array}
\right) \;\; ,
\label{7}
\end{equation}
where $\stL \stR$ mixing is proportional to a large value of the
top-quark mass and is not small. Diagonalizing matrix (\ref{7}) we obtain the
following eigenstates:
\begin{equation}
\left\{
\begin{array}{l}
\tilde{t}_1 =  \cu \stL + \su \stR \\
\tilde{t}_2 = - \su \stL + \cu \stR \;\; ,
\end{array}
\right.
\label{8}
\end{equation}
where $\cu \equiv \cos\thetaLR$, $\su \equiv \sin\thetaLR$, $\thetaLR$ being $\tilde{t}_L
\tilde{t}_R$ mixing angle, and
\begin{equation}
\tan^2\thetaLR =\frac{m_1^2 - \mstL^2}{\mstL^2 -m_2^2} \;\; ,
\;\; m_1^2 \geq \mstL^2 \geq m_2^2 \;\; ,
\label{9}
\end{equation}
where $m_1$ and $m_2$ are the mass eigenvalues:
\begin{equation}
m_{1,2}^2 =
\frac{\mstL^2 + \mstR^2}{2} \pm
\frac{| \mstL^2 - \mstR^2|}{2}
\sqrt{1+\frac{4 \mt^2 A'^{2}_{\mbox{\rm t}}}{( \mstL^2 - \mstR^2)^2}}
\;\; .
\label{10}
\end{equation}

The following relation between $\mstL^2$ and $\msbL^2$
takes place:
\begin{equation}
\mstL^2 = \msbL^2 + \mt^2 + \mZ^2 \cos(2\beta) c_W^2 \;\; ,
\label{11}
\end{equation}
where $\cW^2 \equiv \cos^2\thetaW = 0.77$
($\thetaW$ is the electroweak mixing angle) and $\tan \beta$ is equal to
the ratio of the vacuum averages of two Higgs neutrals, introduced in SUSY
models. Relation (\ref{11}) is central for the present paper; it demonstrates
a large breaking of SU(2) symmetry in the third generation of squarks. The only
hypothesis that is behind this relation is that the main origin of the large
breaking of this SU(2) is in the quark Higgs interaction.

The enhanced electroweak radiative corrections induced by squarks of
 the third generation
depend on 3 parameters: $m_1$, $m_2$ and $\msbL$.
The dependence on the angle $\beta$ is very moderate and, in numerical
 fits, we will
use the rather popular value of $\tan \beta =2$. In what follows
 we will write $ \msb $
 instead of $\msbL$, bearing in mind that the
$\sbR$ squark does not contribute to the corrections under
investigation (let us
note that $\sbL \sbR$ mixing is proportional to $\msb$ and can
be neglected).

Let us present the formulas from \cite{7}, which describe
the enhanced SUSY
corrections to the functions $V_i$:

\begin{equation}
\dSUSYLR \VA = \frac{1}{\mZ^2} [\cu^2 g(m_1, \msb)
+ \su^2 g(m_2, \msb) - \cu^2 \su^2 g(m_1, m_2)] \;\; ,
\label{12}
\end{equation}

\begin{equation}
\dSUSYLR \VR = \dSUSYLR \VA + \frac{1}{3} \YL
\biggl[ \cu^2 \ln \biggl( \frac{m_1^2}{ \msb^2 } \biggr)
     + \su^2 \ln \biggl( \frac{m_2^2}{ \msb^2 } \biggr) \biggr] -
\frac{1}{3} \cu^2 \su^2
h(m_1, m_2) \;\; ,
\label{13}
\end{equation}

\begin{eqnarray}
\dSUSYLR \Vm &=& \dSUSYLR \VA + \frac{2}{3} \YL
s^2 \biggl[ \cu^2 \ln \biggl(\frac{m_1^2}{ \msb^2 } \biggr)
+ \su^2 \ln \biggl( \frac{m_2^2}{ \msb^2 } \biggr) \biggr] + \nonumber \\
&+& \frac{c^2 -s^2}{3}
[\cu^2 h(m_1, \msb) +
 \su^2 h(m_2, \msb)] - \frac{\cu^2 \su^2}{3} h(m_1, m_2)
\;\;
\label{14}
\end{eqnarray}
where
\begin{equation}
g(m_1, m_2) = m_1^2 +m_2^2 -2\frac{m_1^2 m_2^2}{m_1^2 - m_2^2} \ln
\biggl(\frac{m_1^2}{m_2^2} \biggr) \;\; ,
\label{15}
\end{equation}

\begin{equation}
h(m_1, m_2) = -\frac{5}{3} +
\frac{4m_1^2 m_2^2}{(m_1^2 -m_2^2)^2} + \nonumber \\
 \frac{(m_1^2 +m_2^2)(m_1^4 -4m_1^2 m_2^2 +m_2^4)}{(m_1^2 -m_2^2)^3}
\ln \biggl(\frac{m_1^2}{m_2^2} \biggr) \;\; ,
\label{16}
\end{equation}
and $\YL$ is the left-doublet hypercharge,
 $\YL = Q_{\mbox{\rm t}} +Q_{\mbox{\rm b}} =1/3$.

Now that we have all the necessary formulas at our disposal,
 let us start with analysing
the data in the simplest case of the absence of $\stL \stR$
mixing, $\sin \thetaLR =0$. In this case the corrections $\dSUSY V_i$
depend on parameter $\msb$ only, and this dependence is shown in
Figs. 1 - 3 of paper \cite{7}. For $\msb \la 300$ GeV the
theoretical values of all 3 functions $V_i$ become larger than
the experimental
values. Let us begin our fit by taking the value of the Higgs boson mass
 as a free
parameter.
\begin{table}[hbt]
\begin{center}
\begin{tabular}{|c|c|c|c|}
\hline
$\msb$ (GeV) & $\mH$ (GeV) & $\alphashat$ & $\chi^2/ \ndf$ \\
\hline
 & & & \\
100 & $850^{+286}_{-320}$ & $0.113 \pm 0.003$ & 20.3/14 \\
 & & & \\
150 & $484^{+364}_{-235}$ & $0.116 \pm 0.003$ & 18.1/14  \\
 & & & \\
200 & $280^{+240}_{-144}$ & $0.117 \pm 0.003$ & 17.3/14 \\
 & & & \\
300 & $152^{+145}_{-87}$ & $0.118 \pm 0.003$ & 16.3/14 \\
 & & & \\
400 & $113^{+115}_{-68}$ & $0.119 \pm 0.003$ & 15.8/14 \\
 & & & \\
1000 & $77^{+87}_{-47}$ & $0.119 \pm 0.003$ & 15.2/14 \\
 & & & \\
\hline
\end{tabular}
\end{center}
\caption{\label{Tab1}
 Fit of the precision data with SUSY corrections taken into account
in the
case of the absence of $\stL \stR$ mixing, $\sin\thetaLR =0$ and
$\mH$ taken as a free parameter. For $\msb > 300$ GeV, SUSY
corrections become negligible and the SM fit of the data is reproduced.}
\end{table}
In this fit, the mass of the Higgs boson becomes
larger than its Standard Model fit value (\ref{1}). The results of the
fit are
shown in Table~\ref{Tab1}. (To reduce the number of parameters we take $\msg
= \msb$ in this paper. Let us stress that light quarkinos
with masses of the order of $100 - 200$ GeV
are usually allowed only if gluinos
are heavy, $\msg \ge 500$ GeV \cite{12}. In the case of the heavy gluino
correction $\Delta_1$ (eq. (\ref{6})) becomes power-suppressed and we return
to the Standard Model fit value of $\alphashat=0.119$.)
We see how in the SUSY extension of the Standard Model with light
superpartners the fit gets worse.
However, in SUSY the mass of the Higgs boson is no longer a free parameter.
Of the 3 neutral Higgs bosons, the lightest should have a mass less than
 approximately 120 GeV.
 If other Higgs bosons are considerably heavier, this
lightest boson has the same couplings to gauge bosons as in the Standard Model,
so formulas for the Standard Model radiative corrections can be used since
deviations are suppressed as $(\mh / \mA)^2$ ($\mA$ being the mass of the
 heaviest
Higgs).
\begin{table}[hbt]
\begin{center}
\begin{tabular}{|c|c|c|}
\hline
$m_{\tilde{b}}$ (GeV) & $\alphashat$ & $\chi^2/ \ndf$ \\
\hline
100 & $0.110 \pm 0.003$ & 30.2/15\\
150 & $0.115 \pm 0.003$ & 21.9/15 \\
200 & $0.116 \pm 0.003$ & 18.6/15  \\
300 & $0.118 \pm 0.003$ & 16.4/15  \\
400 & $0.119 \pm 0.003$ & 15.8/15 \\
1000 & $0.119 \pm 0.003$ & 15.5/15 \\
\hline
\end{tabular}
\end{center}
\caption{\label{Tab2}
 The same as Table 1, but with a value of the lightest Higgs-boson
 mass
$\mh = 120$ GeV that is about the maximum allowed value in the simplest SUSY
models.}
\end{table}
For the maximal allowed value $\mh = 120$ GeV, the results of the
fit are
shown in Table~\ref{Tab2}. In what follows, we will always take $\mh=120$ GeV
since, for $90$ GeV $< \mh < 120 $ GeV, the results of the fit are practically
the same. This table demonstrates that superpartners should be heavy
if we want to have a good-quality fit of the data.
The next step is to take into account $\stL \stR$ mixing.
\begin{figure}[hbt]
\epsfysize=160pt
\epsfbox{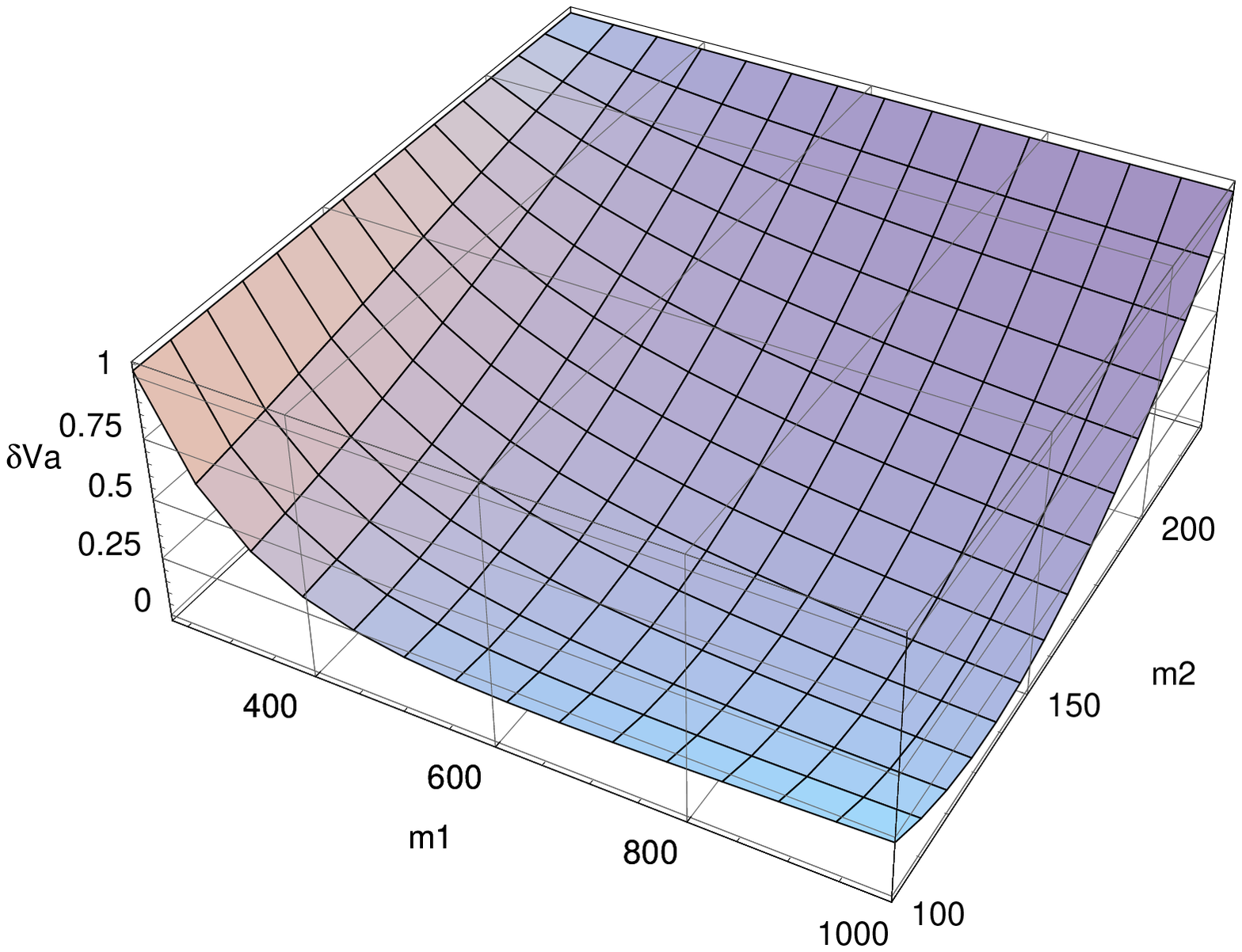}
\epsfysize=160pt
\epsfbox{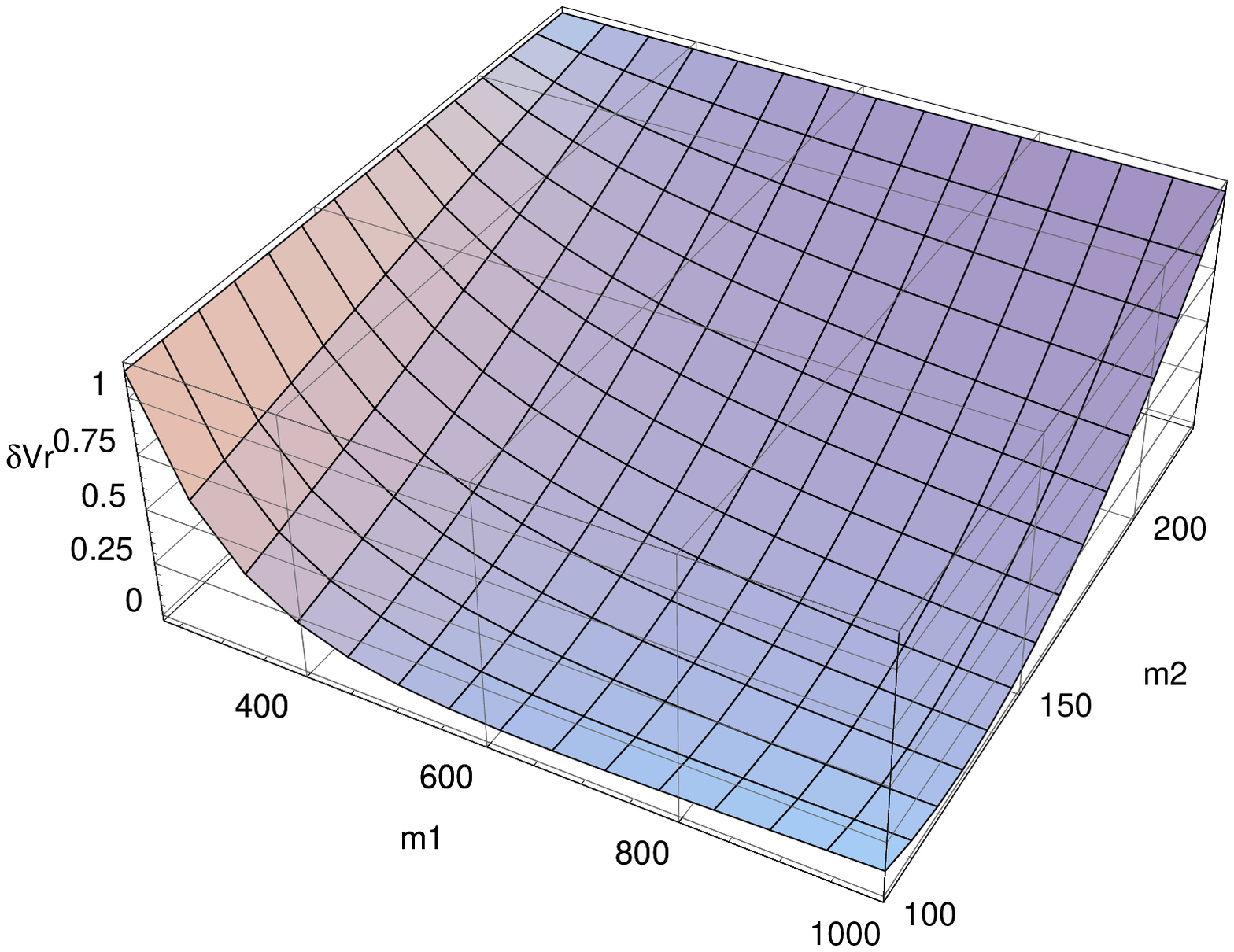}
\epsfysize=160pt
\epsfbox{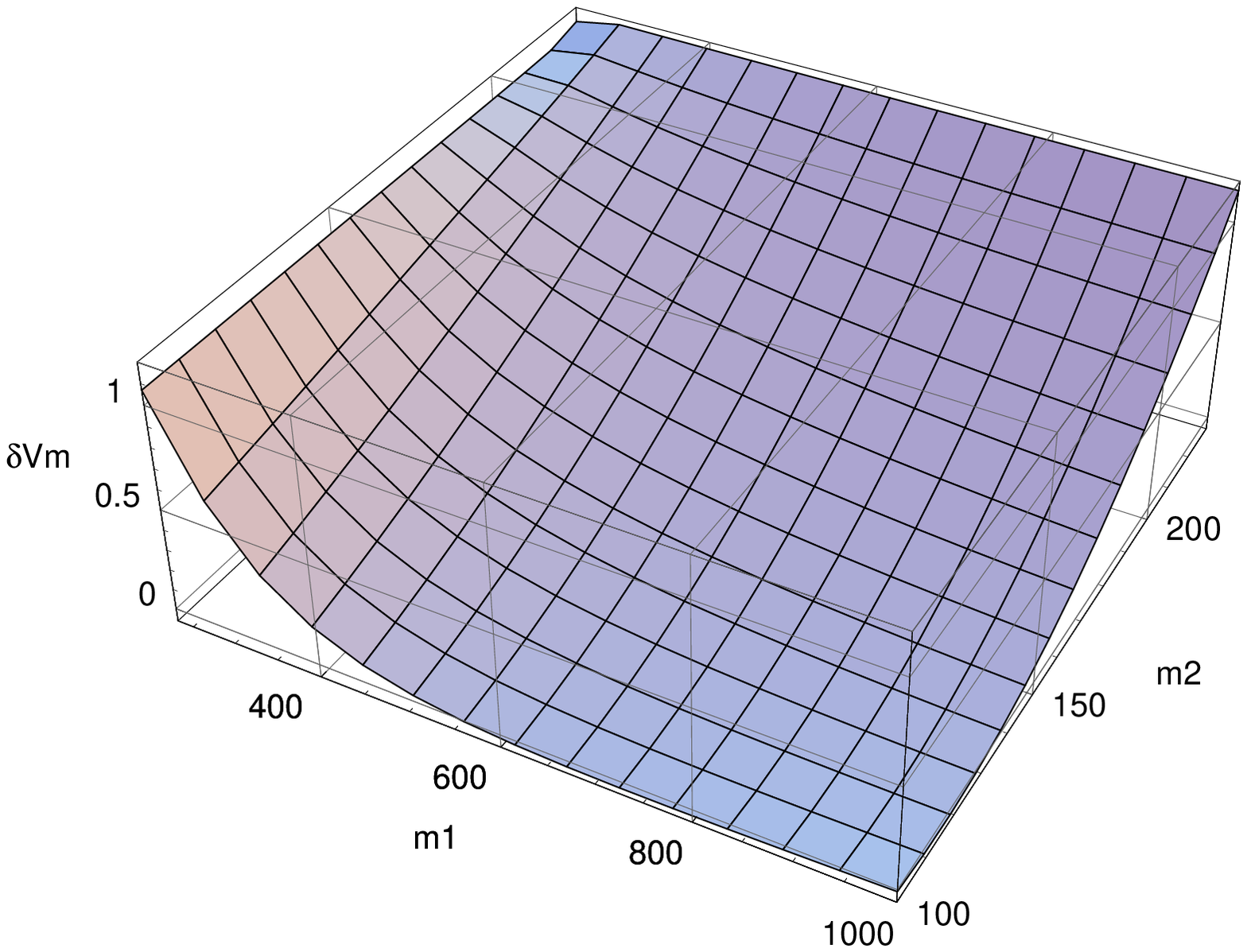}
\caption{\label{Vimsb150}
Values of the $\delta \VA$, $\delta \VR$ and $\delta \Vm$ at
 $\msb = 150$ GeV}
\end{figure}

\begin{figure}[hbt]
\epsfysize=160pt
\epsfbox{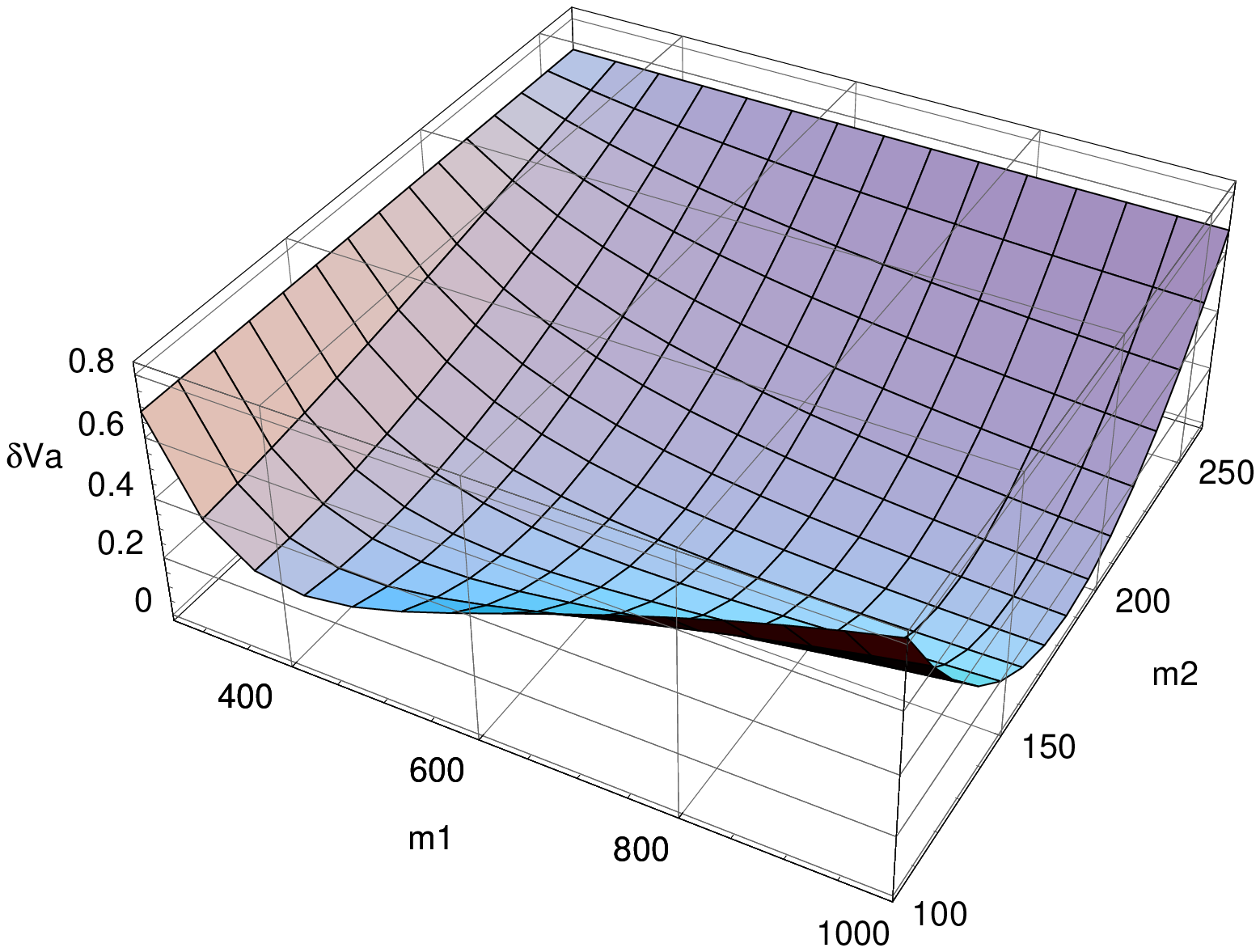}
\epsfysize=160pt
\epsfbox{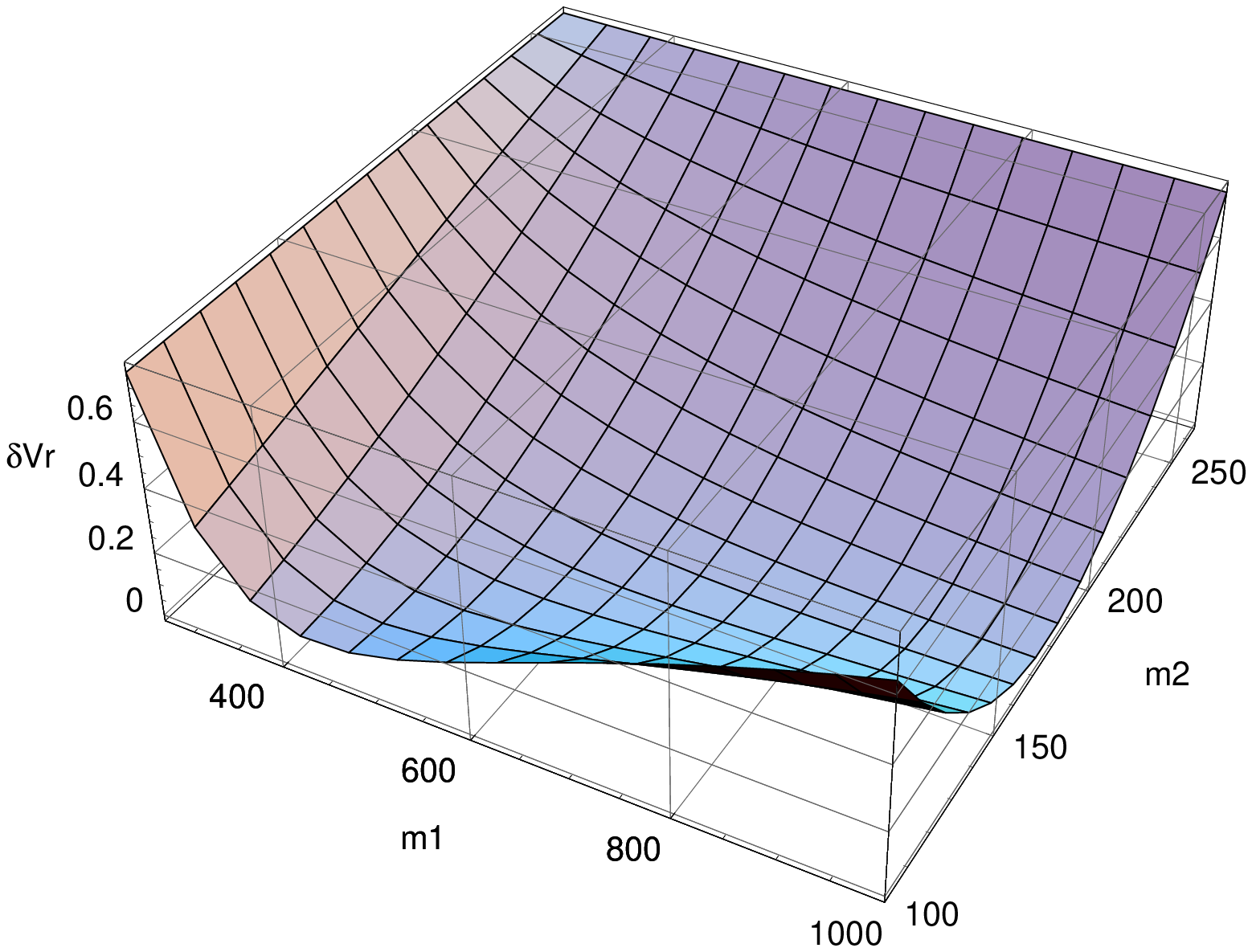}
\epsfysize=160pt
\epsfbox{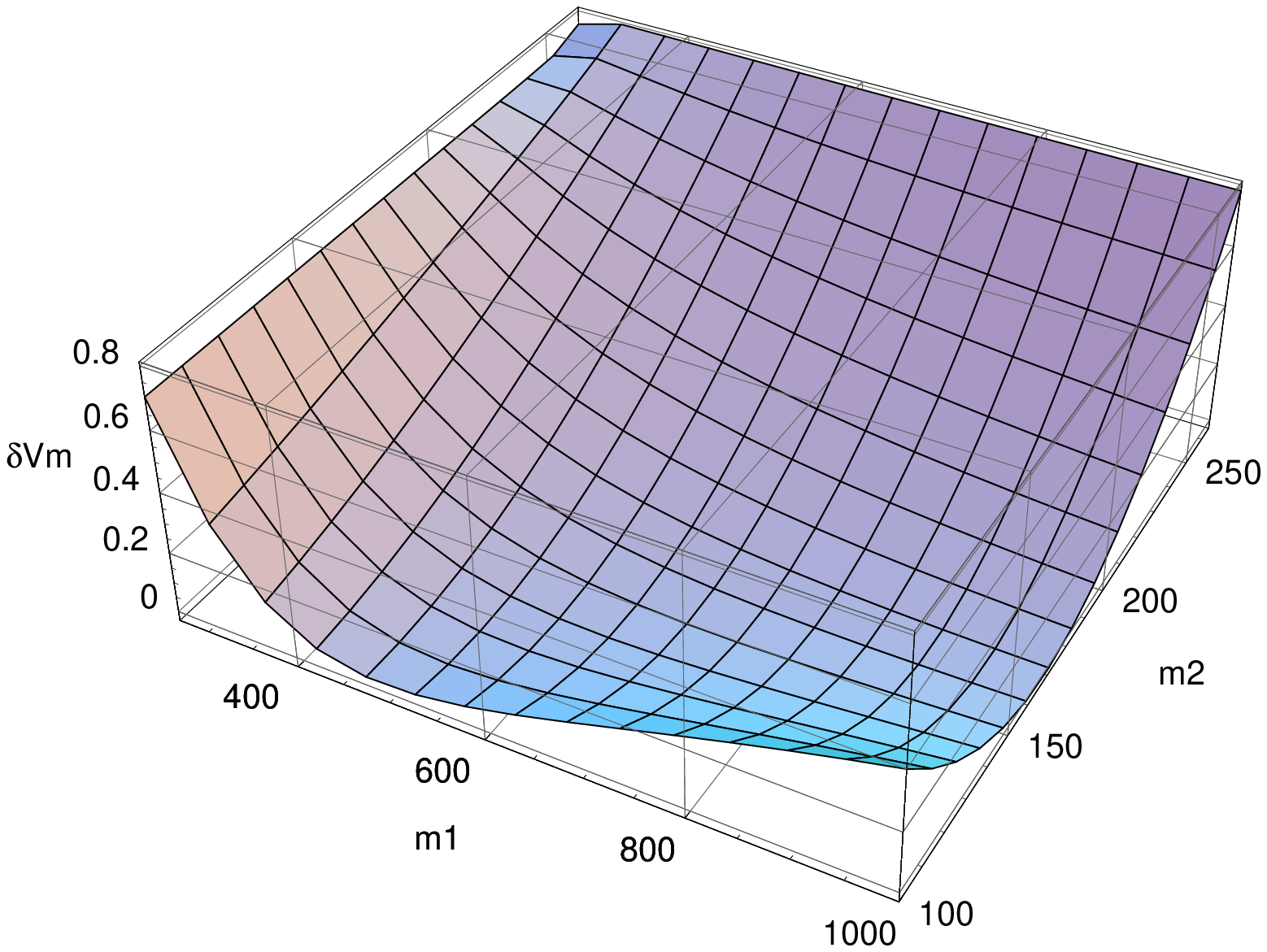}
\caption{\label{Vimsb200}
Values of the   $\delta \VA$, $\delta \VR$ and $\delta \Vm$ at
 $\msb = 200$ GeV}
\end{figure}

In Fig.~\ref{Vimsb150} we show the dependence of SUSY corrections
$\dSUSY V_i$ on $m_1$ and $m_2$ for $\msb = 150$ GeV;
the same in Fig.~\ref{Vimsb200} for
$\msb = 200$ GeV. One clearly sees from these figures that
there exists, even for
such low values of $\msb$, domain of low $m_2$ values
where the enhanced radiative corrections are damped.
 In Fig.\ref{Vimsb200} there is a
valley where $\dSUSY V_i$ reach the minimum values, which are
considerably smaller than 1. This valley starts at $m_2 \approx
\msb$, $m_1\approx 1000$ GeV and goes to $m_2\approx 100$ GeV, $m_1
\approx 400$ GeV. The smallness of the radiative corrections at the point $m_2
\approx \msb$, $m_1\approx 1000$ GeV can be easily understood: here
$\thetaLR \approx \pi/2$, so in $\dSUSY$ $V_A$ only the term proportional to
$g(m_2, \msb)$ remains. However, for $m_2 = \msb$ this term
equals zero. At this end-point of the valley,
 $\sttwo \approx \stL$, $\stone \approx \stR$,
 so $\mstR^2 \gg \mstL^2$, which prevents
the relation between $\mstR$ and $\mstL$ occurring in a
wide class of models. In these models (e.g. in the MSSM) the left and the
right squark masses are equal at the high energy scale and, renormalizing
them to low energies we have $\mstL^2 > \mstR^2$.
 Almost along the whole valley we have
$\tan^2 \thetaLR >1$, which means that $\mstR^2 > \mstL^2$.
This possibility to suppress radiative
corrections was discussed in \cite{770}.
However, in the vicinity of the end-point $m_1 =300$ GeV, $m_2 =70$ GeV
the value of $\tan^2\thetaLR$ becomes smaller than 1 and
$\mstR^2 < \mstL^2$.
\begin{table}[hbt]
\begin{center}
\begin{tabular}{|c|c|c|}
\hline
$m_1$ (GeV) & $\alphashat$ & $\chi^2/ \ndf$ \\
\hline
482 & 0.116(3) & 16.3/15 \\
743 & 0.117(3) & 15.8/15 \\
1289 & 0.117(3) & 15.6/15 \\
\hline
\end{tabular}
\end{center}
\caption{\label{Tab3}
For light $\msb = 150$ GeV, 
light $m_2 = \msb$ and $\mh=120$ GeV, the quality of the fit can be the same
as in the Standard Model, if $m_1$ is heavy enough.}
\end{table}

In Table~\ref{Tab3} we present the results of the fit, assuming
$\msb = 150$ GeV and $\mh = 120$ GeV, along the line of minimum $\chi^2$,
 which is formed at
$m_2 \approx 150$ GeV. We see that for heavy $m_1$ the quality of the fit is not
worse than in the Standard Model.
\begin{figure}[hbt]
\subfigure[]{
\epsfysize=185pt
\epsfbox{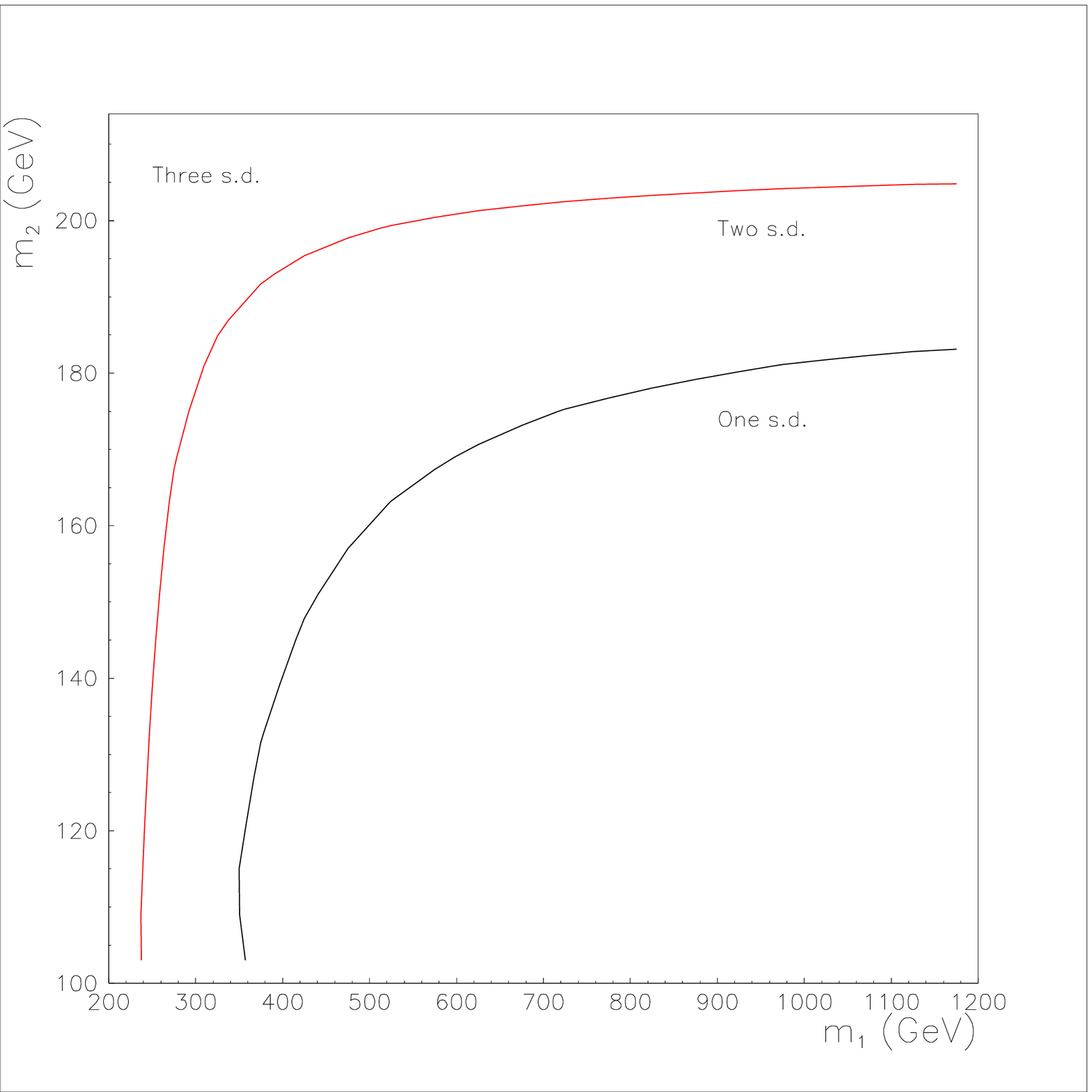}}
\subfigure[]{
\epsfysize=185pt
\epsfbox{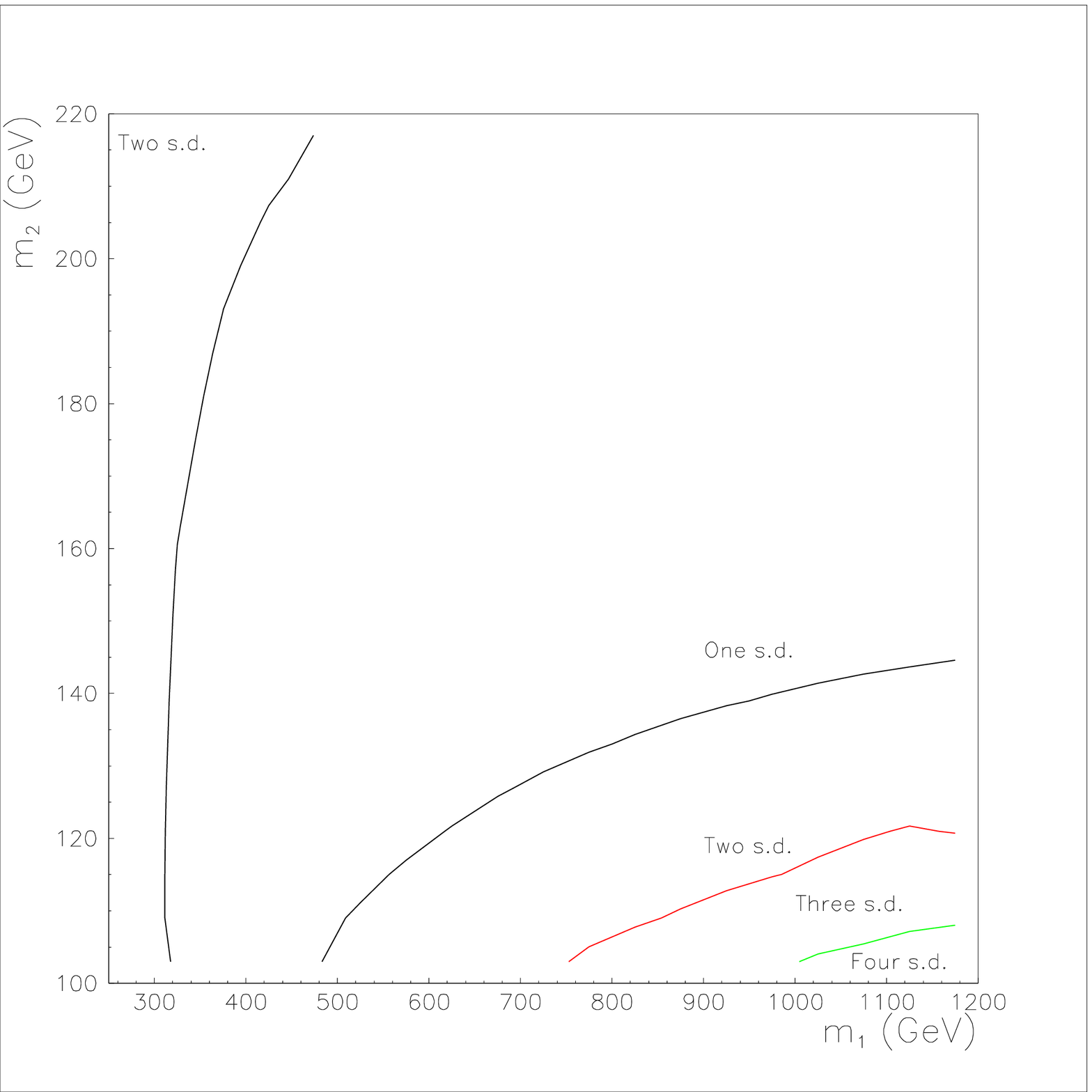}}
\caption{\label{m1m2}
$m_1 - m_2$ exclusion plot for SUSY model, assuming $\mh=120$ GeV,
 $\msg = \msb = \msbL , \tan(\beta)=2$ and eq.(\ref{11}).
Three-parameter fit $(\mt, \alphashat, \alphabar)$ is done to EWWG-99 data.
Figure (a) corresponds to $\msb = 150$ GeV, while (b) corresponds to
$\msb = 200$ GeV.}
\end{figure}
\begin{table}[hbt]
\begin{center}
\begin{tabular}{|c|c|c|c|}
\hline
$m_1$ (GeV) & $m_2$ (GeV) & $\alphashat$ & $\chi^2/ \ndf$ \\
\hline
1296 & 193 & $0.118 \pm 0.003$   & 15.6/15 \\
888 & 167 & $0.118 \pm 0.003$   & 15.8/15 \\
387 & 131 & $0.118 \pm 0.003$   & 16.1/15 \\
296 & 72 & $0.117 \pm 0.003$   & 16.7/15 \\
\hline
\end{tabular}
\end{center}
\caption{\label{Tab4}
For fixed values of $\msb = 200$ GeV and $\mh = 120$ GeV,
 results of the fit along
the valley of minimum $\chi^2$.}
\end{table}
In Table~\ref{Tab4} we show values of $\chi^2$ along its valley of minimum, which
is formed for $\msb = 200$ GeV. Once more we observe that a good
quality of the fit is possible for light superpartners if
 $\stL \stR $ mixing is taken into account.
This effect is clearly seen as a valley of low $\chi^2$ on the plots of
the allowed regions in the plane $m_1 - m_2$ both for $\msb = 150$ GeV 
on Fig. \ref{m1m2}-a
and for $\msb = 200$ GeV on Fig. \ref{m1m2}-b.

The formulas for the enhanced radiative corrections in SUSY extensions of
the Standard Model obtained in \cite{7} were used in the present paper to
fit the data of the
precision measurements of Z-boson decay parameters at LEP and the SLC,
the value of $\mW$ and $\mt$ at the Tevatron.
The fit with SUSY corrections, assuming a small value of $\msb$,
the absence of $\stL \stR$ mixing, and $\mh = 120$ GeV,
leads to the growth of the $\chi^2$ value.
Thanks to the decoupling property of the SUSY
extension in the case of heavy squarks, the results of the Standard Model
 fit are
reproduced.  However, even for comparatively light sbottom and small
mass of one of the two stops, values of $\stL \stR$ mixing can be found
that have
$\dSUSY V_i$ small and $\chi^2$ almost the same as
in the Standard Model.

We are grateful to L.B.~Okun for stimulating discussions.
A.V.~Novikov
is grateful to G.~Fiorentini for his hospitality in Ferrara,
where part of this work was done. A.N.~Rozanov is grateful to
CPPM-IN2P3-CNRS for supporting this work.
The research of
I.G., A.N., V.N. and M.V. is supported by RFBR
grants 96-02-18010, 96-15-96578, 98-02-17372 and 98-02-17453;
that of A.N., V.N. and M.V. by INTAS-RFBR grant 95-05678 as well.
We thanks Susy Vascotto for the kind help in the preparation of the manuscript.

\end{document}